\documentstyle[floats,prd,aps,epsf,eqsecnum,12pt]{revtex}

\makeatletter
\newbox\tempboxa
\newdimen\captionboxsubcount 
\def\capsize#1{\captionboxsubcount=#1pt}
\newdimen\captionboxsub
\captionboxsub=\hsize \advance\captionboxsub by -\captionboxsubcount
\advance\captionboxsub by -\captionboxsubcount
\long\def\@makecaption#1#2{
 \setbox\@tempboxa\hbox{#1 #2}
 \ifdim \wd\@tempboxa >\captionboxsub 
\rightskip=\captionboxsubcount \leftskip=\captionboxsubcount #1 #2 
\else \hbox to\hsize{\hfil\box\@tempboxa\hfil} 
 \fi}
\makeatother
\capsize{30}

\begin{document}

\begin{titlepage}

\begin{flushright}
\begin{minipage}{5cm}
\begin{flushleft}
\small
\baselineskip = 13pt
SU--4240--663\\
hep-ph/9708113 \\
August, 1997
\end{flushleft}
\end{minipage}
\end{flushright}

\begin{center}
\Large\bf
Toy Model for Breaking Super Gauge Theories at the Effective Lagrangian 
Level 
\end{center}

\vfil

\footnotesep = 12pt

\begin{center}
\large
Francesco {\sc Sannino}
\footnote{Address after September 1, 1997: 
{\rm  Dept.~of~Physics,~Yale~University, New Haven, CT 06520-8120.}}
\footnote{
Electronic address : {\tt sannino@suhep.phy.syr.edu}}
\quad and \quad
Joseph {\sc Schechter}
\footnote{
Electronic address : {\tt schechte@suhep.phy.syr.edu}}
\end{center}

\begin{center}
\it 
\qquad
Department of Physics, Syracuse University, 
Syracuse, NY 13244-1130, USA
\end{center}

\vfill

\begin{center}
\bf
Abstract
\end{center}

\begin{abstract}
\baselineskip = 17pt
We propose a toy model to illustrate how the effective Lagrangian for 
super QCD might go over to the one for ordinary QCD by a mechanism 
whereby the gluinos and squarks in the fundamental theory decouple below 
a given supersymmetry breaking scale $m$. The implementation of this approach 
involves a suitable choice of possible supersymmetry breaking terms. An 
amusing feature of the model is the emergence of the ordinary QCD degrees 
of freedom which were hidden in the auxiliary fields of the supersymmetric 
effective Lagrangian. 
\end{abstract}
\begin{flushleft}
\end{flushleft}

\vfill

\end{titlepage}

\setcounter{footnote}{0}

\section{Introduction}
In the last few years there has been a great flurry of interest in the 
effective Lagrangian approach to supersymmetric gauge theories. This was 
stimulated by some papers of Seiberg \cite{Seiberg} and Seiberg and Witten 
\cite{Seiberg-Witten} in which a number of fascinating ``exact results'' were 
obtained. There are already several interesting review articles 
\cite{IntSeiberg,Peskin,DiVecchia}. 

It is natural to hope that information obtained from the more highly 
constrained  supersymmetric gauge theories can be used to learn more 
about ordinary gauge theories, notably QCD.  This is not necessarily as simple 
as it might appear at first. At the fundamental gauge theory level the 
supersymmetric theories contain gluinos and squarks in addition to the
 ordinary gluons and quarks. At the effective supersymmetric Lagrangian level, 
all of the physical fields are composites involving at least one gluino and 
one squark. This means that none of them should appear in an effective 
Lagrangian for ordinary QCD. Where the mesons and glueballs, which are the 
appropriate fields for an effective QCD Lagrangian, actually do appear are in 
the auxiliary fields of the supermultiplets, which get eliminated from the 
theory. For example, the scalar and pseudoscalar glueball fields are hidden 
in the auxiliary field
\begin{equation}
F\sim F_{mn}F^{mn} + i F_{mn}\tilde{F}^{mn} \ , \nonumber 
\end{equation}
(see Eq.~(\ref{F})) while the meson fields are hidden in the auxiliary field
\begin{equation}
\left(F_T\right)_{ij}\sim \psi_i \tilde{\psi}_j \ ,
\end{equation}
(see Eq.~(\ref{FT})). Here ${F}_{mn}$ is the gauge field strength, 
 $\tilde{F}_{mn}$ is its dual while $\psi_{i}$ and $\tilde{\psi}_j$ are 
quarks and anti-quarks. These fields appear in effective Lagrangians 
for super Yang Mills theory and super QCD developed some time ago by several 
authors \cite{refVY,refTVY,refADS}. They were designed to saturate the 
gauge theory anomalies and their features were extensively discussed 
\cite{AmatiReview,RussianReview}.

The simplest approach to relate the supersymmetric effective theories 
to the ordinary ones is to add suitable supersymmetry breaking terms. 
This has been carried out by a number of groups 
\cite{Masiero-Veneziano,MPRV,ASYPY,EHS,ADKM}. The 
standard procedure assumes the breaking terms to be ``soft'' in order 
to keep the theory close to the supersymmetric one. Indications were 
that the soft symmetry breaking was beginning to push the models in the 
direction of the ordinary gauge field cases. However the resulting effective 
Lagrangians were not written in terms of QCD fields.

In this paper we will provide a toy model for expressing the ``completely 
broken'' Lagrangian in terms of the desired ordinary QCD fields. Since 
we will no longer be working close to the supersymmetric theory we 
will not have the protection of supersymmetry for deriving ``exact results''. 
In practice this means a greater arbitrariness in the choice of the 
supersymmetry breaking terms. The advantage of our approach is that we end 
up with an actual QCD effective Lagrangian. 

Our method involves several ideas and assumptions. We will add (not soft) 
supersymmetry breaking pieces to the known \cite{refVY,refTVY,refADS} effective super Lagrangians 
At scales above a value $m$ it will be assumed that supersymmetry is a good 
approximation. In this region the auxiliary $F$ fields should be integrated 
out as usual. At scales below $m$ we imagine supersymmetry to be broken 
so that it is more appropriate to integrate out the fields which are 
composites involving ``heavy'' gluinos and squarks and {\it retain} the 
auxiliary fields. Constraints on the supersymmetry breaking terms will 
be obtained by requiring the trace of the energy momentum tensor in this 
region to agree (at one loop level) with that of ordinary QCD. The two 
regimes will be ``matched'' so that below the scale $m$, the appropriate 
invariant scale is $\Lambda_{QCD}$, while above $m$ the invariant scale 
is that for the SUSY theory (denoted $\Lambda$). It will be noted that 
a reasonable picture results if we take the dominant terms to be those 
obtained by neglecting the K\"ahler terms in the original supersymmetric 
effective Lagrangian. This feature is analogous to Seiberg's 
treatment \cite{Seiberg} of supersymmetric effective Lagrangians with different flavor 
numbers. It results in a dominant piece of the QCD effective Lagrangian 
possessing a kind of tree level holomorphicity. Physically this corresponds 
to the explicit realization of the axial and trace anomalies by the model. 

Our approach should be clarified in section \ref{sec:SUSYYANGMILLS} which 
treats the breaking of supersymmetric Yang Mills down to ordinary Yang Mills. 
In that case the dominant ``holomorphic'' term is sufficient to explain 
the gluon condensation which underlies the ``bag model'' \cite{GJJK} approach 
to confinement. The more complicated case of QCD with $N_f(<N_c)$ massless 
quarks is treated in section \ref{sec:QCD}. In this model the 
``$U_A(1)$ problem'' can be resolved and the need for a ``non holomorphic'' 
addition to the model understood. 

Further improvements and extensions of the present approach are briefly 
discussed in section \ref{sec:Discussion}

\section{From Super Yang Mills to Yang Mills}
\label{sec:SUSYYANGMILLS}
The effective Lagrangian for Super Yang Mills was given \cite{refVY} by Veneziano 
and Yankielowicz (VY) and is described by the Lagrangian 
\begin{equation}
{\cal L} = \frac{9}{\alpha} \int d^2\!\theta d^2\!{\bar{\theta}} 
\left(S S^{\dagger}\right)^{\frac{1}{3}} + 
\left\{\int d^2\theta\,S\left[{\rm ln}\left(\frac{S}{\Lambda^3}\right)^{N_c} - 
N_c \right] + {\rm h.c.} \right\} \ ,
\label{VY}
\end{equation}
 where $\Lambda$ is the super $SU(N_c)$ Yang Mills invariant scale and the 
chiral superfield $S$ stands for\footnote{Notation is identical to that of 
Wess and Bagger \cite{Wess-Bagger}.} the composite object 
 $S=\frac{g^2}{32\pi^2}W^{\alpha}_{a}W_{\alpha a}$. Here $g$ is the gauge 
coupling constant and $W^{\alpha}_{a}$ is the supersymmetric field strength. 
At the component level 
 $S(y)=\phi(y) + \sqrt{2} \theta \psi(y) + \theta^2 F(y)$, where $y^m=x^m + 
i \theta \sigma^m \bar{\theta}$ and 
\begin{eqnarray}
\phi&=&-\frac{g^2}{32\pi^2} \lambda^2 \ , \label{phi}\\
\sqrt{2}\psi&=&\frac{g^2}{32 \pi^2} 
\left[\sigma^{mn} \lambda_a F_{mn,a} - i \lambda_a D_a\right] \ , 
\label{psi} \\
F&=&\frac{g^2}{32\pi^2} \left[ -\frac{1}{2} F^{mn}_a F_{mn,a} - 
\frac{i}{4} \epsilon_{mnrs}F^{mn}_a F^{rs}_ a \right. \nonumber \\
&&\left.+ D_a D_a - i \bar{\lambda}_{a}\bar{\sigma}^{m} 
\stackrel{\leftrightarrow}{\nabla} \lambda_a + i \partial^m J_m^5 \right] 
\label{F} \ .
\end{eqnarray}
Here $\lambda^{\alpha}_a$ is the gluino field, $F_{mn,a}$ the gauge field 
strength, $D_a$ the auxiliary field for the gauge multiplet and 
$J^5_m = \bar{\lambda}_a \bar{\sigma}_m \lambda_a$ is the axial current. 

We interpret the complex field $\phi$ as representing scalar and 
pseudoscalar gluino balls while $\psi$ is their fermionic partner. 
The auxiliary field $F$, which gets eliminated in the supersymmetric 
context, actually is seen to contain scalar and pseudoscalar glueball 
type objects of interest in the ordinary Yang Mills theory. 
It should be noted that the VY model is not an effective 
Lagrangian in the same sense as the chiral Lagrangian of pions which 
describes light degrees of freedom and can therefore be systematically 
improved \cite{WGL} by the introduction of higher derivative terms. Rather, the 
physical particles in the VY model are heavy. Nevertheless the model 
has been considered to be a very instructive one. It describes the vacuum 
of the theory and, if the particles which appear are taken to be the 
important physical ones, a model for their interactions which saturates 
the anomalous Ward identities at tree level. These anomalies arise 
in the axial current of the gluino field, the trace of the energy momentum 
tensor and in the special superconformal current. In supersymmetry these 
three anomalies belong to the same supermultiplet 
\cite{Ferrara-Zumino} and hence 
are not independent. For example
\begin{eqnarray}
\theta^m_m&=&3\,N_c\,\left(F + F^{*}\right) = 
-\frac{3 N_c g^2}{32\pi^2} F^{mn}_a F_{mn,a} \ , \label{trace}\\
\partial^m J_{m}^5 &=&2i\, N_c \, \left(F-F^{*}\right)= 
\frac{N_c g^2}{32\pi^2} \epsilon_{mnrs}F^{mn}_a F^{rs}_a \ ,
\label{axial}
\end{eqnarray}
where we employed the classical equations of motion to get 
the second equalities on each line. Note that $\theta^m_m$ is 
normalized so that, for any theory, 
$\displaystyle{\frac{1}{4}\langle0|\theta^m_m|0\rangle}$ is the vacuum energy density.  

The effective Lagrangian at tree level is well known \cite{refVY} to yield 
gluino condensation of the form
\begin{equation}
\langle\phi\rangle=-\frac{g^2}{32\pi^2} \langle\lambda^2\rangle = 
\Lambda^3 e^{\frac{2\pi i k}{N_c}} \ ,
\label{susyvacuum}
\end{equation}
where $k=0,1,2,\cdots,(N_c -1)$. The presence of the integer $k$ indicates 
$N_c$ different equivalent vacuum solutions (Witten index \cite{Witten}) and 
arises from the multi--valuedness\footnote{A formulation of the model 
which provides a single valued Lagrangian has recently been proposed by 
Kovner and Shifman \cite{Kovner-Shifman}.} 
of the logarithm in Eq.~(\ref{VY}). Super 
Yang Mills possesses a discrete $Z_{2N_c}$ symmetry which, by choosing 
a given vacuum, is spontaneously broken to $Z_2$ 
($\lambda \rightarrow -\lambda$). 

It is a widespread hope that the information available on the Super Yang 
Mills theory can be transferred in some way to the ordinary Yang Mills case. 
The most straightforward approach is to add a ``soft'' supersymmetry breaking 
term to the Lagrangian. This was carried out by Masiero and Veneziano 
\cite{Masiero-Veneziano} 
who introduced a ``gluino mass term'' in the Lagrangian
\begin{equation}
{\cal L}=\cdots + m \left(\phi + \phi^*\right) \ ,
\label{soft}
\end{equation}
with the softness restriction $m\ll \Lambda$. The results 
\cite{Masiero-Veneziano} of this model indicate that 
the theory is ``trying'' to approach the ordinary Yang Mills case: The 
spin $0$ and spin $1/2$ particles split from each other and their masses 
each pick up a piece linear in m. One of the $N_c$ different vacua becomes 
the true minimum. Furthermore the vacuum value of the glueball field 
$\langle F \rangle$ is no longer zero (as required by supersymmetry) but picks up 
a piece proportional to $m$. (Actually they eliminate $F$ in the usual way.) 
In ordinary Yang Mills theory a non zero value of 
$\langle F+F^* \rangle$ is associated 
with confinement. 

It seems very desirable to extend this model to the case of large 
$m(\gg\Lambda)$ in which the superparticles actually decouple from the 
theory and the theory gets reexpressed in terms of ordinary glueball 
fields. Clearly this is a difficult non--perturbative problem. Here 
we propose a toy model which accomplishes these goals. 
Our approach is based on the following three assumptions:
\begin{itemize}
\item[i)]{As in Seiberg's analysis \cite{Seiberg} of the super QCD models with 
varying number of flavors we shall concentrate completely on the 
superpotential 
$\displaystyle
{W=S\left[{\rm ln}\left(\frac{S}{\Lambda^3}\right)^{N_c} - N_c\right]}$. This 
contains all the information on the anomaly structure and seems to be 
the least model dependent part of the effective Lagrangian.}
\item[ii)]{We will show that the generalization of the supersymmetry breaking 
term Eq.~(\ref{soft}) to 
\begin{equation}
{\cal L}=\cdots + m^{\delta} \phi^{\gamma} + {\rm h.c.} \ ,
\label{breaking}
\end{equation}
where $\delta=4-3\gamma$ and 
$\displaystyle{\gamma=\frac{12}{11}}$ automatically accomplishes 
the decoupling of the underlying gluino degree of freedom at the scale $m$. 
The deviation of the exponent $\gamma$ from unity is being thought of as 
an effective description of the evolution of the symmetry breaker 
Eq.~(\ref{soft}) for large $m$.}
\item[iii)]{In supersymmetry, the field $F$ is eliminated by its equation 
of motion $\displaystyle{\frac{\partial V}{\partial F}=0}$. 
Since the Yang Mills 
fields of interest are contained in $F$ we shall adopt an alternative 
procedure in which the heavy gluino ball field $\phi$ is eliminated 
by its equation of motion $\displaystyle{\frac{\partial V}{\partial \phi}=0}$. 
In the present case we will not include contributions to $V$ from 
the K\"ahler terms. We will see that this leads to a reasonable picture. 
Intuitively it seems natural to neglect the kinetic terms for fields whose 
 masses increase enough so that they can be 
integrated out.}
\end{itemize}

The potential of our model 
\begin{equation}
V(F,\phi)=-F\,{\rm ln}\left(\frac{\phi}{\Lambda^3}\right)^{N_c} - 
m^{\delta} \phi^{\gamma} + {\rm h.c.} \ ,
\label{totpotentialVY}
\end{equation}
provides the equation of motion, 
 $\displaystyle{\frac{\partial V}{\partial \phi}}=0$ for eliminating $\phi$:
\begin{equation}
\phi^{\gamma}=-\frac{N_c F}{\gamma m^{\delta}}  \ .
\label{eomVY}
\end{equation}
Our physical requirement is that the presence of the symmetry breaker 
Eq.~(\ref{breaking}) should convert the anomalous quantity $\theta^m_m$ 
into the appropriate one for the ordinary Yang Mills theory. 
This is in the same spirit as the well known \cite{Wittendec} criterion for 
decoupling a heavy flavor (at the one loop level) in QCD. 
We compute 
$\theta^m_m$ at tree level from the formula \cite{schechter}: 
\begin{equation}
\theta^m_m=4V - \left[4F\frac{\partial V}{\partial F} 
+ 3 \phi \frac{\partial V}{\partial \phi} + {\rm h.c.}\right] \ ,
\label{stVY}
\end{equation}
which takes the dimensions of the fields $F$ and $\phi$ into account. 
Using Eq.~(\ref{eomVY}) we obtain
\begin{equation}
\theta^m_m=\frac{4N_c}{\gamma}\left(F + F^*\right)= -
\frac{4N_c}{\gamma} \left(\frac{g^2}{32\pi^2}F^{mn}_aF_{mn,a}\right) \ .
\label{stYM}
\end{equation}
Now the 1--loop anomaly in the underlying theory is given by
\begin{equation}
\theta^m_m=-b\frac{g^2}{32\pi^2}F^{mn}_aF_{mn,a} \ ,
\label{stunder}
\end{equation}
where $b=3N_c$ for supersymmetric Yang Mills and $b=\frac{11}{3}N_c$ for 
ordinary Yang Mills. In order that Eq.~(\ref{stYM}) match Eq.~(\ref{stunder}) 
for ordinary Yang Mills we evidently require 
 $\displaystyle{\gamma=\frac{12}{11}}$ as mentioned above. With $\phi$ 
eliminated in terms of $F$ the potential becomes
\begin{equation}
V(F)=-\frac{11N_c}{12}F\left[{\rm ln}\left(\frac{-11 N_c F}
{12 m^{\frac{8}{11}} 
\Lambda^{\frac{36}{11}}}\right) - 1 \right] + {\rm h.c.} \ .
\label{YM}
\end{equation}
We now check that this is consistent with a physical picture in which 
the gauge coupling constant evolves according to the super Yang Mills 
beta--function above scale $m$ and according to the Yang Mills beta--function 
below scale $m$. Since the coupling constant at scale $\mu$ is given 
by $\displaystyle{\left(\frac{\Lambda}{\mu}\right)^b = 
\exp\left(\frac{-8\pi^2}{g^2(\mu)}\right)}$, the matching at $\mu=m$ requires 
$\displaystyle{\left(\frac{\Lambda}{m}\right)^b=
\left(\frac{\Lambda_{YM}}{m}\right)^{b_{YM}}}$, which yields
\begin{equation}
\Lambda_{YM}^4=m^{\frac{8}{11}} \Lambda^{\frac{36}{11}} \ ,
\label{matchVY}
\end{equation}
in agreement with the combination appearing in Eq.~(\ref{YM}).
The lagrangian in Eq.~(\ref{YM}) manifestly depends only on quantities 
associated with the Yang Mills theory, the gluino degree of freedom 
having been consistently decoupled. Equation~(\ref{YM}) thus seems to be 
a reasonable candidate for the potential term of a model describing 
the trace anomaly in Yang Mills theory. 

The model is seen to contain 
both a scalar glueball field ${\rm Re}F$ and a pseudoscalar glueball field 
${\rm Im}F$. In addition to discussing the vacuum structure, one might  imagine 
adding suitable scale invariant kinetic terms to upgrade the model 
to describing physical fields. However there is a non--trivial 
feature present. To see this let us investigate the potential in more 
detail. The vacuum solutions are obtained from the equations
\begin{eqnarray}
\frac{\partial V}{ \partial {\rm Re}F} &=& -\frac{11N_c}{6}\, 
{\rm ln}\left(\frac{11N_c}{12}\frac{|F|}{\Lambda^4_{YM}}\right) = 0 \ , 
\label{ReF}\\
\frac{\partial V}{ \partial {\rm Im}F} &=& -i\frac{11N_c}{12} 
\, {\rm ln}\left(\frac{F}{F^*}\right) = 0 \ . \label{ImF}
\end{eqnarray}
(The effects of a non zero vacuum angle, $\theta$ will be discussed later). 
Satisfying Eq.~(\ref{ReF}) and Eq.~(\ref{ImF}) requires 
$\displaystyle{N_c\langle F \rangle={\rm real}= -\frac{12}{11}\Lambda^4_{YM}}$; the 
sign has been chosen for consistency with Eq.~(\ref{YM}). The reality 
of $\langle F\rangle$ also follows from parity invariance. {}For the second derivatives 
we have 
\begin{equation}
\left\langle \frac{\partial^2 V}{\partial\left({\rm Re}F\right)^2}\right\rangle 
= -\left\langle \frac{\partial^2 V}{\partial\left({\rm ImF}\right)^2}\right\rangle 
=\frac{2}{\Lambda_{YM}^4}\left(\frac{11 N_c}{12}\right)^2 \ .
\label{2V}
\end{equation}
This shows that, if we interpret both ${\rm Re}F$ and ${\rm Im}F$ 
as physical degrees 
of freedom, the vacuum solution obtained above has an instability associated 
with fluctuations in the ${\rm Im}F$ direction. In other language, while 
${\rm Re}F$ has 
a positive mass squared coefficient ${\rm Im}F$ has a wrong sign mass squared 
coefficient. At first glance, this would seem to be a serious deficiency 
of the model. However, in earlier discussions \cite{joe} 
of the $U_A(1)$ problem 
($\eta^{\prime}$ mass problem) in the framework of a toy Lagrangian 
which exactly mocks up the $U_A(1)$ anomaly, it was found necessary 
to {\it postulate}  a wrong sign mass squared term for the pseudoscalar 
glueball field in order to achieve a non zero $\eta^{\prime}$  mass. 
No kinetic term for the pseudoscalar was to be written so its equation 
of motion relates it, in fact, to the $\eta^{\prime}$ field. (This 
mechanism will be illustrated in the next section when quarks are 
included in the model.) From this point of view, the prediction of a wrong 
sign mass squared term for ${\rm Im}F$ is a welcome feature.  

The above discussion suggests that, in the present case, we should 
also eliminate ${\rm Im}F$ by its equation of motion Eq.~(\ref{ImF}). The 
solution is clearly ${\rm Im}F=0$. Substituting this back into Eq.~(\ref{YM}) and 
using the notation 
\begin{equation} 
H=\frac{11N_c }{3}\frac{g^2}{32\pi^2}F^{mn}_aF_{mn,a} \ ,
\label{Hdef}
\end{equation}
leads to the potential function
\begin{equation}
V(H)=\frac{H}{4}\, {\rm ln} \left(\frac{H}{8e\Lambda^4_{YM}}\right) \ .
\label{YMH}
\end{equation}
This may be considered as a zeroth order model 
\cite{schechter,MS,SST} 
for Yang Mills theory 
in which the only field present is a scalar glueball. $V(H)$ has 
a minimum at $\langle H\rangle = 8\Lambda^4_{YM}$, at which point 
$\langle V\rangle=-2 \Lambda^4_{YM}$. {}From Eq.~(\ref{Hdef}) this is 
seen to correspond to a magnetic--type condensation of the glueball 
field $H$. The negative sign of $\langle V\rangle$ is consistent 
with the bag model \cite{GJJK} 
in which a ``bubble'' with $\langle V \rangle = 0$ 
is stabilized against collapse by the zero point motion of the particles 
within. A number of phenomenological questions have been discussed 
using toy models based on Eq.~(\ref{YMH}) \cite{SST,GJJS,GJJS2}.  

It is also interesting to discuss the dependence of the 
Lagrangian on the QCD 
vacuum angle $\theta$. The potential in Eq.~(\ref{totpotentialVY}) gets 
modified to 
\begin{equation}
V(F,\phi)=-F\left[{\rm ln}\left(\frac{\phi}{\Lambda^3}\right)^{N_c} - 
i\left(\theta + 2\pi k\right)\right] - m^{\delta}\phi^{\gamma} + {\rm h.c.} \ ,
\label{thetapot}
\end{equation} 
where we have now displayed the arbitrary integer $k$ reflecting 
the multi-valued 
nature of the logarithm. Since the ``mass-type'' term 
$\displaystyle{m^{\delta}\phi^{\gamma}}$ is present it is not possible 
to rotate $\theta$ away. Integrating out $\phi$, as above yields
\begin{equation}
V(F)=-\frac{N_cF}{\gamma} 
\left[{\rm ln} \left(\frac{-N_c F}{\gamma \Lambda^4_{YM}}\right) - 1 
-i\frac{\gamma \hat{\theta}}{N_c}\right] + {\rm h.c.} \ ,
\label{YMTheta}
\end{equation}
where $\displaystyle{\hat{\theta}=\left(\theta + 2\pi k\right)}$. Note that 
in this framework periodicity, corresponding to the transformation  
 $\displaystyle{\theta \rightarrow \theta + 2\pi n}$ for integer $n$, is 
maintained by choosing a different branch of the logarithm according to 
 $k\rightarrow k-n$. It is also possible to see that the $N_c-$fold degeneracy 
of vacuum states present in super Yang Mills is broken so that a 
vacuum with a particular value of $k$ has minimum energy; this is 
discussed in Appendix A. 

We have seen that the complex scalar gluino ball field $\phi$ can be 
integrated out and its degrees of freedom transferred to the ordinary 
glueball variables. It remains to check that its super partner, $\psi$ 
in Eq.~(\ref{psi}), suitably decouples in the present model. This 
is easy to see since its mass is proportional to 
$\displaystyle{\left\langle \frac{\partial^2 W}{\partial\phi^2}\right\rangle 
=\frac{N_c}{\langle \phi\rangle}}$. This may be rewritten, 
using Eq.~(\ref{eomVY}) and 
$\displaystyle{\langle F \rangle = - \frac{12 \Lambda^4_{YM}}{11 N_c}}$, as:
\begin{equation}  
m_{\psi}\propto \frac{N_c}{\Lambda^{\frac{11}{3}}_{YM}}m^{\frac{2}{3}} \ .
\label{psi-mass}
\end{equation}
The constant of proportionality depends on the choice of $\psi$ kinetic 
term but reasonable choices can be seen not to change our conclusion. 
It is seen that $m_{\psi} \rightarrow \infty$ in the case that $\Lambda_{YM}$ 
remains fixed and $m\rightarrow \infty$; thus $m_{\psi}$ decouples. 

It is interesting to note that the potential for the ordinary Yang Mills 
theory in Eq.~(\ref{YM}) or Eq.~(\ref{YMTheta}) also displays a 
tree-level holomorphic 
structure. This is due to our assumption that the effect of the K\"ahler 
term, which would give an $F^* F$ term in $V$, is negligible for 
the decoupled theory. The effects of possible non holomorphic terms 
can calculated as (presumably small) perturbations.

\section{From Super QCD to QCD.}
\label{sec:QCD}

The next step is clearly the addition of $N_f$ flavors of zero mass quark 
superfields to the underlying super Yang Mills theory. Our goal 
is to see how the decoupling of superpartners discussed in the previous 
section might get generalized to this more complicated case. {}For simplicity 
we will restrict attention to $N_c \neq 2$ (to avoid a special extra symmetry) 
and $N_f<N_c$ (to avoid extra relevant composite baryonic superfields 
in the effective Lagrangian). The needed ``mesonic'' composite 
superfield is the complex $N_f \times N_f$ matrix 
\begin{equation}
T_{ij}= Q_i \tilde{Q}_j \ ,
\label{T: def}
\end{equation}
where $i$ and $j$ are flavor indices and the quark and anti-quark superfields 
are expanded as $Q=\phi_{Q} + \sqrt{2}\theta \psi_{Q} + \theta^2 F_{Q}$ and 
$\tilde{Q}=\phi_{\tilde{Q}} + \sqrt{2}\theta \psi_{\tilde{Q}} + 
\theta^2 F_{\tilde{Q}}$.  
The matrix $T$ has the decomposition
 $T=Q\tilde{Q}=t + \sqrt{2} \theta \psi_{T} + \theta^2 F_T$ with 
 \begin{eqnarray}
 t&=&\phi_{Q}\phi_{\tilde{Q}} \ , \label{tpiccolo}\\
 \psi_T&=&\psi_{Q}\phi_{\tilde{Q}} + \phi_{Q}\psi_{\tilde{Q}} \ , \\
 F_T&=& \phi_{Q}F_{\tilde{Q}} +  
 F_{Q}\phi_{\tilde{Q}} - \psi_Q\psi_{\tilde{Q}} \ ,
\label{FT} 
\end{eqnarray}
where flavor indices are not shown.
 We consider here the complex field $t$ as representing scalar and 
pseudoscalar squark-antisquark composites while $\psi_T$ is 
their fermionic partner. The auxiliary field $F_T$ contains the scalar 
and pseudoscalar ordinary QCD mesons ($\psi_{Q}\psi_{\tilde{Q}}$). 
This suggests that, as for Yang Mills, the 
auxiliary matrix field $F_T$ can be regarded, once the super-partners are 
decoupled, as the actual QCD meson variable.   

The effective QCD superpotential for the present case was 
given in Ref.~\cite{refTVY} by 
Taylor, Veneziano and Yankielowicz (TVY):  
\begin{equation}   
W_{TVY}=S \left[{\rm ln} 
\left(\frac{S^{N_c - N_f} {\rm det} T}{\Lambda^{3N_c - N_f}}\right) 
- \left(N_c - N_f\right)
\right] \ .
\label{TVY}
\end{equation}
This form saturates at tree level the SUSY QCD anomalies; it is 
invariant under a well-known $U_R(1)$ axial transformation which corresponds 
to a particular linear combination of the quark and gluino axial 
transformations. For convenience, some needed results on the symmetry 
structure are briefly summarized in Appendix~\ref{app:b}. 
In most applications of this model it is reasonable to focus on the 
``light'' degrees of freedoms in $T$ and integrate out the ``heavy'' degrees 
of freedom in $S$ by using 
 $\displaystyle{\frac{\partial W_{TVY}}{\partial S}=0}$. This leads 
to $\displaystyle{S^{N_c-N_f}=\frac{\Lambda^{3N_c - N_f}}{{\rm det}T}}$ which, 
on substituting for $S$ back in Eq.~(\ref{TVY}), yields the Affleck Dine 
Seiberg (ADS) \cite{refADS} model:
\begin{equation}
W_{ADS}=-\left(N_c - N_f\right)\left[\frac{\Lambda^{3N_c-N_f}}
{{\rm det} T}\right]^{\frac{1}{N_c - N_f}}\ .
\label{ADS}
\end{equation}
Both Eq.~(\ref{TVY}) and Eq.~(\ref{ADS}) yield potentials whose minima 
correspond to ``runaway vacua'', i.e. $\langle T\rangle \rightarrow \infty$. 
This is interpreted as an inconsistency of super QCD with massless quarks 
when $N_f<N_c$. Although this behavior is very different from what is 
expected for ordinary QCD it does properly match on \cite{Seiberg} 
to the $N_f=0$ and $N_f=N_c$ cases.

The straightforward approach of adding a ``soft'' supersymmetry breaking 
term to the super QCD effective Lagrangian was discussed in 
Ref.~\cite{Masiero-Veneziano} and more recently by Aharony et al. 
\cite{ASYPY}. These authors have used a breaking term of the type
\begin{equation}
{\cal L}=\cdots - \rho {\rm Tr}\left[t^{\dagger}t\right] \ ,
\label{abreak}
\end{equation}
where $\rho$ is a positive constant and the field $t$ was defined in 
Eq.~(\ref{tpiccolo}). Note that this term is invariant under the full 
chiral $U_L(N_f)\times U_R(N_f)$ group. It was observed that the resulting 
theory was starting to behave like QCD in the sense that the squark 
condensate $\langle t\rangle$ decreased while a non-zero quark condensate 
$\langle F_T\rangle$ appeared. Now we would like to generalize the discussion 
of the breaking of the VY model given in the last section to the TVY case. 
Again we will restrict attention to the 
superpotential and integrate out the gluino and squark composite objects 
 $\phi$ and $t$ in favor of the gluon and quark composite fields $F$ 
and $F_T$.  Suitable new symmetry breaking terms will be added so that 
both the squark as well as the gluino underlying degrees of freedom 
will decouple below the (for simplicity) single scale $m$. This will 
be implemented by requiring the trace anomaly at scales greater than 
$m$ to agree with that of SUSY QCD and the trace anomaly at scales less 
than $m$ to agree with that of ordinary QCD. Furthermore we preserve 
the quark fields' axial $U_A(1)$ anomaly while transferring its realization 
from $t$ to $F_T$. 

In the previous VY case the potential in Eq.~(\ref{totpotentialVY}) of our 
broken model could be called ``holomorphic'' in the sense that it 
involved complex fields (generically $\chi_i$) and had the structure 
${\cal F}\left(\chi_i\right)+h.c.$. This form of ``holomorphicity'' arises 
from the process of realizing the anomalies in both the supersymmetric 
as well as the ordinary QCD. In the supersymmetric case this kind 
of holomorphicity is eventually lost due to the $FF^{*}$ type pieces 
in the K\"ahler terms. In turn this guarantees the positive definiteness 
of the potential. In ordinary QCD, however, the positiveness of the potential 
is actually undesirable as noted in the discussion below Eq.~(\ref{YMH}). 
Hence it is natural to expect the ``holomorphic'' terms to play a dominant 
role in the ordinary QCD potential. This causes the feature (as we have 
already seen in Eq.~(\ref{2V})) that for any single field $\chi$
\begin{equation}
\frac{\partial^2 V}{\left(\partial {\rm Im}\chi\right)^2 }=-
\frac{\partial^2 V}{\left(\partial {\rm Re}\chi\right)^2} \ .
\label{instability}
\end{equation}
In the VY case this led to a wrong sign pseudoscalar glueball mass term. 
However this was noted to actually be needed for solving the $U_A(1)$ 
problem; the solution involved integrating out the appropriate field 
by its equation of motion. We will employ a similar procedure in the 
present TVY case. The holomorphic form of the potential that we will, at 
first, achieve will be seen to have some desirable features. However 
it will lead to an unphysical value of the vacuum energy density. We will 
show that this problem may be cured by the addition of a suitable 
non holomorphic piece (as a perturbation) which, however, does not 
affect the anomaly structure of the theory. Our first thought, and one to 
which we will return in the future, about the choice of symmetry breakers
 to be added for the TVY case is to take the sum of 
Eq.~(\ref{breaking}) and a 
simple modification of Eq.~(\ref{abreak})  to $-\rho^{a}\left[{\rm Tr} \left(
t^{\dagger}t\right)\right]^b$. However this form leads to complications 
in the attempt to explicitly eliminate the fields $\phi$ and $t$ from the 
potential by their equations of motion. For this initial analysis of our 
approximation scheme we will choose supersymmetry breaking terms which 
make the elimination of $\phi$ and $t$ as simple as possible. We shall, 
however, require that all symmetry breaking terms we add preserve the 
$U_A(1)$ anomaly of QCD; the job of satisfying the anomaly 
is assigned to the log term which the model inherits from its supersymmetric 
parent.  
    
The potential of the unbroken model from Eq.~(\ref{TVY}) is  
\begin{equation}  
V_0(F,\phi;F_T,t)=-F\,{\rm ln} \left(\frac{\phi^{(N_c -N_f)}{\rm det}(t)}
{\Lambda^{(3N_c - N_f)}}\right) - \phi {\rm Tr}\left[F_Tt^{-1}\right] +
{\rm h.c.} \ .
\label{tvypot}
\end{equation}  
Here $\Lambda$ is the supersymmetric QCD invariant scale. 
Let us rewrite the log in such a way that the as yet unrelated 
conventional QCD scale, $\Lambda_{QCD}$ is displayed: 
\begin{equation}
-F\,{\rm ln} \left(\frac{\phi^{N_c}}
{\Lambda^{3N_c}}\right) -F\,{\rm ln} \left(\frac{{\rm det}F_T}{\Lambda_{QCD}^{3N_f}}
\right) -  
F\,{\rm ln} \left(\frac{\left(\Lambda_{QCD}^3\Lambda\right)^{N_f}{\rm det}(t)}
{{\rm det}F_T\phi^{N_f}}\right) \ . 
\label{tvyY}
\end{equation}
Note that the second term in Eq.~(\ref{tvyY}) saturates the QCD $U_A(1)$ 
anomaly. 
It is now convenient to define the following composite operator
\begin{equation}
Y^{N_f}\equiv \frac{\phi^{N_f}{\rm det}F_T}{{\rm det}(t)} \ .
\label{Y}
\end{equation}
It is easy to see, since $F_T$ transforms in the same way as $t$ under 
global chiral transformations that $Y$ is invariant 
under the matter field $U_A(1)$ transformation, and hence will not affect 
the QCD $U_A(1)$ anomaly. 
For convenience we will consider, rather than $t$ and $\phi$, 
 $Y$ and $\phi$ as the variables to be integrated out. 
We will consider these fields to decouple at the single scale $m$.   
It is worth noticing that the last term in Eq.~(\ref{tvypot}) is holomorphic, 
scale invariant and invariant with respect to the full global chiral group. 
{}For simplicity we will cancel it in the supersymmetry breaking potential:  
\begin{equation}
V_{SB}=+\phi {\rm Tr}\left[F_T t^{-1}\right] - m^{\delta}\phi^{\gamma} 
+ m^{\tilde{\delta}} Y^{\tilde{\gamma}} + h.c. \ ,
\label{sbtvy}
\end{equation}
where $\delta=4-3\gamma$ and $\tilde{\delta}=4-4\tilde{\gamma}$. 
The complete model potential is initially taken as the ``holomorphic'' 
structure:
\begin{eqnarray}
V(F,\phi;F_T,Y)&=&-F\,{\rm ln} \left(\frac{\phi^{N_c}}
{\Lambda^{3N_c}}\right) -\
F\,{\rm ln} \left(\frac{{\rm det}F_T}{\Lambda_{QCD}^{3N_f}}
\right) + N_f\,F\,{\rm ln} \left(\frac{Y}{\Lambda_{QCD}^3 \Lambda}\right) 
\nonumber \\   
&&{}-m^{\delta}\phi^{\gamma} + m^{\tilde{\delta}}Y^{\tilde{\gamma}} 
+{\rm h.c.} \ ,
\label{breakingSQCD}
\end{eqnarray}
The breaking potential displayed in Eq.~(\ref{breakingSQCD}) is a  
generalization of one used in the super 
Yang Mills case (Eq.~(\ref{breaking})). 
 The equations of motion 
 $\displaystyle{\frac{\partial V}{\partial \phi}=0}$ and 
$\displaystyle{\frac{\partial V}{\partial Y}=0}$ for the unwanted 
degrees of freedom now take the very simple forms
\begin{eqnarray}
\phi^{\gamma}&=&-\frac{N_c F}{\gamma m^{\delta}} \ , \nonumber \\
Y^{\tilde{\gamma}}&=&-\frac{N_f F}{\tilde{\gamma}
m^{\tilde{\delta}}} \ .
\label{phiYequations}
\end{eqnarray}
Eliminating $\phi$ and $Y$ in terms of $F$ yields:
\begin{eqnarray}
V(F,F_T)&=&-\left(\frac{N_c}{\gamma}-\frac{N_f}{\tilde{\gamma}}\right)\, F 
\left[{\rm ln}\left(\frac{-N_c F}{\gamma m^{\delta}\Lambda^{3\gamma}}\right) 
- 1\right] - F\, {\rm ln} \left(\frac{{\rm det}F_T}{\Lambda_{QCD}^{3N_f}}\right)
\nonumber \\
&&{}+\frac{N_f}{\tilde{\gamma}}F\,{\rm ln}
\left(\frac{N_f \gamma m^{\delta} \Lambda^{3\gamma}}
{N_c \tilde{\gamma} m^{\tilde{\delta}}
\Lambda_{QCD}^{3\tilde{\gamma}}\Lambda^{\tilde{\gamma}}}\right) + 
{\rm h.c.}  \ .
\label{QCDprepot}
\end{eqnarray}
{}As in the Yang Mills case, we require that 
the presence of the supersymmetry breaking terms in
 Eq.~(\ref{breakingSQCD}) convert the trace of the energy momentum tensor
 $\theta^m_m$ into 
the appropriate one for the QCD theory. At the tree level for the 
effective lagrangian in Eq.~(\ref{breakingSQCD}), $\theta^m_m$ may be 
evaluated \cite{schechter} as: 
\begin{equation}
\theta^m_m=4 V - \left[4F\, \frac{\partial V}{\partial F} 
+ 3\phi\frac{\partial V}{\partial \phi} + 4 Y\frac{\partial V}{\partial Y} 
+ 3 {\rm Tr}\left[F_T \frac{\partial V}{\partial F_T} \right] + {\rm h.c.}
\right] \ ,
\label{traceSQCD}
\end{equation}
which, by using Eq.~(\ref{phiYequations}), yields
\begin{equation} 
\theta^m_m=
\left[4\left(\frac{N_c}{\gamma}-\frac{N_f}{\tilde{\gamma}}\right) 
+ 3N_f\right] \left(F + F^*\right) \ .
\label{stQCD}
\end{equation}
If we set $N_f=0$, the previous formula correctly reproduces the pure 
Yang Mills $\theta^m_m$ (Eq.~(\ref{stYM})). The 1--loop trace anomaly in the 
underlying theory is given in Eq.~(\ref{stunder}), where now 
$\displaystyle{b=3N_c - N_f}$ 
for SUSY QCD and $\displaystyle{b=\frac{11}{3}N_c - \frac{2}{3}N_f}$ for 
ordinary QCD. To match Eq.~(\ref{stQCD}) with the underlying theory we 
require:
\begin{equation}
4\left(\frac{N_c}{\gamma}-\frac{N_f}{\tilde{\gamma}}\right) 
+ 3N_f=\frac{11}{3}N_c - \frac{2}{3}N_f \ .
\label{matchSQCD}
\end{equation}
To completely fix the two unknowns, $\gamma$ and $\tilde{\gamma}$, we need 
another relation. According to our physical picture we require that 
the gauge coupling 
evolves according to the super QCD beta-function above scale $m$ 
and according to the QCD 
beta-function below scale $m$. On physical grounds the QCD effective 
potential in 
Eq.~(\ref{QCDprepot}) should depend only on the 
QCD invariant scale $\Lambda_{QCD}$. 
Hence we will uniquely fix $\gamma$ by imposing 
\begin{equation}
m^{\delta}\Lambda^{3\gamma}=\Lambda_{QCD}^{4} \ .
\label{gammamat}
\end{equation}
At the 1-loop level  (to be consistent with 1-loop trace anomaly matching) 
the constraint in Eq.~(\ref{gammamat}) provides
\begin{equation}
\gamma=\frac{4}{3}\frac{b}{b_{QCD}}=\frac{12 N_c - 4 N_f}{11 N_c - 2 N_f} \ , 
\label{gammaQCD}
\end{equation}
which for $N_f=0$ is consistent with the Yang Mills determination  
of $\gamma$ deduced in the previous section.  Substituting 
Eq.~(\ref{gammaQCD}) into Eq.~(\ref{matchSQCD}) we obtain
\begin{equation}
\tilde{\gamma}=\frac{4b}{b_{QCD}+3b}=12\frac{3N_c - N_f}{38N_c - 11 N_f} 
\ .
\label{tgammaQCD}
\end{equation}
As a check of decoupling the quantity  
$\displaystyle{
\frac{m^{\delta}\Lambda^{3\gamma}}
{m^{\tilde{\delta}} \Lambda_{QCD}^{3\tilde{\gamma}}
\Lambda^{\tilde{\gamma}}}}$ becomes independent of the gluino 
mass scale and the original SUSY QCD scale and in fact equal to one. Then 
 Eq.~(\ref{QCDprepot}) may be simply written as:
\begin{eqnarray}
V(F,F_T)&=&-\left(\frac{N_c}{\gamma}-\frac{N_f}{\tilde{\gamma}}\right)\, F 
\left[{\rm ln}\left(\frac{-N_c F}{\gamma \Lambda_{QCD}^{4}}\right) 
- 1\right] - F\, {\rm ln} \left(\frac{{\rm det}F_T}{\Lambda_{QCD}^{3N_f}}\right)
\nonumber \\
&&{}+\frac{N_f}{\tilde{\gamma}}F\,{\rm ln}
\left(\frac{N_f \gamma}
{N_c \tilde{\gamma}}\right) + 
{\rm h.c.}  \ .
\label{QCDprepotr}
\end{eqnarray}
where $\gamma$ and $\tilde{\gamma}$ are given in 
Eq.~(\ref{gammaQCD}) and Eq.~(\ref{tgammaQCD}). 
This expression involves just the composite 
fields which are made from quarks and gluons; $F$ contains a scalar and 
a pseudoscalar glueball field while the matrix $F_T$ contains $N_f^2$ scalar 
meson fields and $N_f^2$ pseudoscalar meson fields.  $V(F,F_T)$ has the 
nice features that it satisfies the QCD anomalies, has the ``holomorphic'' 
structure and displays at the effective Lagrangian level, the appropriate 
decoupling of the squark and gluino degrees of freedom with the correct 
one loop matching condition at the breaking scale $m$. We will now see that 
this potential appears to solve the $U_A(1)$ problem by generating an 
$\eta^{\prime}$ mass term. As discussed in the previous section we consider 
that no kinetic term for the field ${\rm Im}F$ should be added to our model 
so that its equation of motion simply becomes
\begin{equation}
0=\frac{\partial V}{\partial{\rm Im}F}=
-i\left(\frac{N_c}{\gamma} - \frac{N_f}{\tilde{\gamma}}\right)\, 
{\rm ln}\,\left(\frac{F}{F^*}\right) 
- i \,{\rm ln}\,\left(\frac{{\rm det} F_T}{{\rm det} F_T^{\dagger}}\right) \ .
\label{ImFint}
\end{equation}
Introducing $F=-|F|e^{i\Phi}$ and ${\rm det}F_T=|{\rm det}F_T|e^{i\Phi_T}$ 
then gives the relation:
\begin{equation}
\Phi=-\frac{\Phi_T}{\left(\frac{N_c}{\gamma} - 
\frac{N_f}{\tilde{\gamma}}\right)} \ .
\label{PhiPhiT}
\end{equation}
Now rewriting Eq.~(\ref{QCDprepotr}) 
with the help of Eq.~(\ref{ImFint}) yields 
\begin{eqnarray}
V=&-&2\left(\frac{N_c}{\gamma} - 
\frac{N_f}{\tilde{\gamma}}\right){\rm Re}F\,
\left[{\rm ln}\,\left(\frac{|{\rm Re}F|N_c}{e\gamma \Lambda^4_{QCD}}\right)-
{\rm ln}\,|\cos \Phi|\right] \nonumber \\
~&+&2\frac{N_f}{\tilde{\gamma}}\,{\rm Re}F\, {\rm ln}\,
\left(\frac{N_f \gamma}{N_c \tilde{\gamma}}\right) 
-{\rm Re}F\,\ln\,\left(\frac{{\rm det}\left(F_T^{\dagger} F_T\right)}
{\Lambda^{6N_f}_{QCD}}\right) \ . 
\label{ReFV}
\end{eqnarray}
Note that the $\eta^{\prime}$ field (pseudoscalar meson singlet) is 
proportional to the phase $\Phi_T$. Expanding the 
$\displaystyle{{\rm ln}\,|\cos \Phi|}$ term to second order in $\Phi$ and 
using Eq.~(\ref{PhiPhiT}) yields the $\Phi_T$ dependence;
\begin{equation}
V=-\frac{{\rm Re}F}{\left(\frac{N_c}{\gamma} - 
\frac{N_f}{\tilde{\gamma}}\right)} \Phi_T^2 + \cdots \ , 
\label{etapm}
\end{equation}
which when we replace $-{\rm Re}F$ by $\langle - {\rm Re}F\rangle$ and 
accept the sign choice found in the broken VY model case 
(after Eq.~(\ref{ImF})) amounts to a correct sign mass term for the 
$\eta^{\prime}$. Thus the potential in Eq.~(\ref{QCDprepotr}) seems to 
know, in a fairly detailed way, something about QCD. 

\begin{figure}[htbp]
\begin{center}
\epsfxsize=250pt
\ \epsfbox{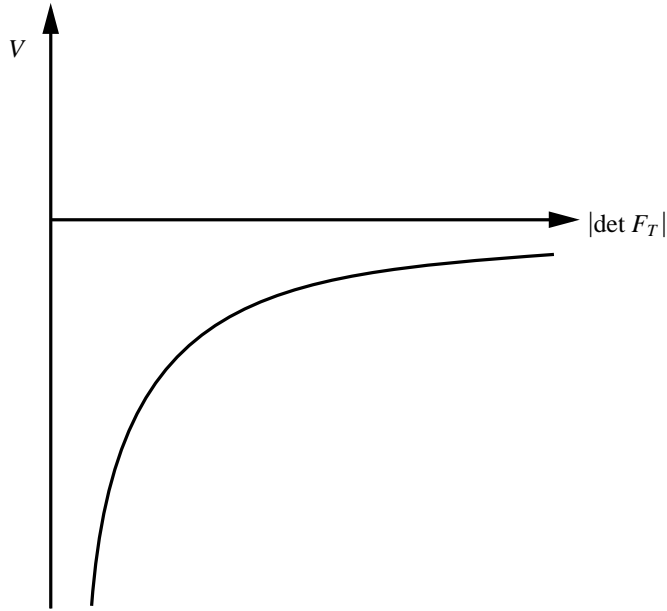}
\end{center}
\caption[]{Sketch of the potential $V$ in Eq.~(\ref{VQads}).}
\label{adsb}
\end{figure}

However, a difficulty arises when we look for a stable minimum of the 
potential (\ref{QCDprepotr}). We see that 
\begin{equation}
\frac{\partial V}{\partial |{\rm det}F_T|}=-\frac{2{\rm Re}F}{|{\rm det} F_T|} \ ,
\label{FTder}
\end{equation}
which, noting again that $\langle - {\rm Re}F\rangle$ is expected to be 
positive, leads to a minimum of the potential when $|{\rm det}F_T|=0$, 
which in turns corresponds to $\langle V\rangle \rightarrow - \infty$. 
One possibility for solving this problem might be to eliminate ${\rm Re}F$ 
in favor of $|{\rm det}F_T|$. Since we have already eliminated ${\rm Im}F$ 
in favor of $\Phi_T$, this amounts to completely eliminating the ``heavy'' 
glueball degrees of freedom in favor of the ``light'' mesonic degrees 
of freedom. Such a procedure corresponds, at the supersymmetric level, 
to going from the TVY to the ADS model. In any event the equation for 
integrating out ${\rm Re}F$ is obtained from Eq.~(\ref{QCDprepotr}) as:
\begin{equation} 
0=\frac{\partial V}{\partial {\rm Re}F}=-2
\left(\frac{N_c}{\gamma} - 
\frac{N_f}{\tilde{\gamma}}\right)\,{\rm ln}\,
\left(\frac{N_c|F|}{\gamma \Lambda_{QCD}^4}\right) - 
2\,{\rm ln}\, \left(\frac{|{\rm det}F_T|}{\Lambda_{QCD}^{3N_f}}\right)+
2\frac{N_f}{\tilde{\gamma}}\,{\rm ln}\left(\frac{N_f \gamma}
{N_c \tilde{\gamma}}\right) \ .
\label{ReFint}
\end{equation}
Substituting this back into $V$ yields the expression:
\begin{equation}
V=-2\frac{p\gamma \Lambda^4_{QCD}}{N_c} 
\left(\frac{N_f \gamma}{N_c \tilde{\gamma}}\right)^
{\frac{N_f}{\tilde{\gamma}p}} \,{|\cos \Phi|}\,
{\left|\frac
{\Lambda_{QCD}^{3N_f}}{{\rm det}F_T} \right|^{\frac{1}{p}}} \ ,
\label{VQads}
\end{equation}
with 
$\displaystyle{p=\frac{N_c}{\gamma} - \frac{N_f}{\tilde{\gamma}}
=\frac{11}{12}\left(N_c - N_f \right)}$. A sketch of this 
expression is shown in Fig.~\ref{adsb}.
It exhibits ``fall to the origin'' 
(like the classical hydrogen atom $s$-wave state) as already 
indicated\footnote{It is amusing to note that the present potential 
behaves qualitatively like the negative of the runaway potential for the 
ADS model in the supersymmetric limit.} 
by Eq.~(\ref{FTder}). Hence an additional ingredient seems required in the 
model. We will take the new ingredient to be a ``non holomorphic'' 
supersymmetry breaker. It will be required, as for the breaking terms in 
Eq.~(\ref{sbtvy}), to respect the 
full chiral group. {}For simplicity it will be 
treated as a perturbation in the sense that Eq.~(\ref{ImFint}) and 
Eq.~(\ref{ReFint}) will still be assumed to hold. In order not to disturb 
the matching  conditions based on Eq.~(\ref{traceSQCD}) we will assume the term to be 
scale invariant. A suitable form is 
\begin{equation}
\delta V_{SB}=A \left(F^{*}F\right)^{r}\, 
\left[{\rm Tr}\left(F^{\dagger}_T F_T\right)\right]^s \ ,
\label{deltasb}
\end{equation}
where $A$ is a positive constant and $8r+6s=4$. {}For simplicity we will 
now consider the $N_f=1$ case. The results can be easily generalized to 
the case $N_f>1$. Using Eq.~(\ref{ReFint}) for eliminating $|F|$ 
and substituting back into Eq.~(\ref{deltasb}) we have
\begin{equation} 
\delta V_{SB}=\frac{B}{|F_T|^d} \ ,
\label{sbft}
\end{equation}
where the coefficients $B$ and $d$ are 
\begin{eqnarray}
B&=&A\left[\frac{\gamma}{N_c} \Lambda^{4+\frac{3}{p}}_{QCD} 
\left(\frac{\gamma}{N_c \tilde{\gamma}}\right)^
{\frac{1}{p\tilde{\gamma}}}\right]^{2r} \ ,
\label{capB} \\
d&=&\frac{1}{p} - s \left(\frac{3}{2p}+2\right). 
\label{ld}
\end{eqnarray}
The new potential, obtained by adding together 
Eq.~(\ref{VQads}) and Eq.~(\ref{sbft}) 
has a stable minimum when we choose $s< 0$. 
The minimum condition also forces a non zero value 
for the quark-antiquark condensate $\langle F_T \rangle$ 
and no spontaneous breaking 
of strong CP symmetry ($\langle \Phi_T \rangle =0$).
The generalization of this approach for $N_f>1$ would lead to a 
spontaneous breaking of chiral symmetry. It thus seems that a model 
of the present type, which includes ``holomorphic'' terms to satisfy the 
anomalies and supersymmetry breaking terms which include a 
``non holomorphic'' piece can give a description of the properties of the 
QCD vacuum in general agreement with our expectations \cite{smrst}.

\section{Discussion}
\label{sec:Discussion}
We have shown that it is possible to construct toy models which explicitly 
realize the breaking of super gauge theories down to ordinary gauge 
theories in the effective Lagrangian framework. The resulting Lagrangians 
are built from the ordinary QCD glueball and meson fields. An important 
feature is the realization of the axial anomaly and the trace anomaly both 
at the supersymmetric as well as at the ordinary levels. The price we have 
paid for getting explicit results far away from the supersymmetric situation 
is the introduction of supersymmetry breaking terms chosen especially to 
simplify the calculation, while preserving the required symmetry properties. 

{}For orientation we first studied the supersymmetric Yang Mills theory 
(described by the VY Lagrangian) and its breaking to ordinary Yang Mills. 
We learned, perhaps somewhat surprisingly, that both the decoupling of the 
gluino in the underlying theory as well as the expected nature of the 
Yang Mills vacuum could be understood with the neglect of the original 
K\"ahler terms. Our procedure led to a non supersymmetric potential 
(Eq.~(\ref{YM})) which possessed a kind of tree-level holomorphicity. It would 
be interesting to systematically investigate non holomorphic corrections 
due to the K\"ahler terms. 

We next investigated the breaking of supersymmetric QCD (described by the 
TVY Lagrangian) down to ordinary QCD. {}Following a procedure similar 
to the previous one led to a plausible model for QCD which was consistent 
with the spontaneous breakdown of chiral symmetry and the generation 
of an $\eta^{\prime}$ mass ($U_A(1)$ problem). In this more complicated 
case the holomorphic piece of the potential, Eq.~(\ref{QCDprepotr}) had some nice 
features but gave a vacuum energy (see Fig.~\ref{adsb}) unbounded from below. 
This was cured by the addition of a suitable non holomorphic piece, 
Eq.~(\ref{deltasb}). 
{}For the QCD situation, both the K\"ahler terms as well 
as a supersymmetry breaking term like in Eq.~(\ref{abreak}) 
can give non holomorphic 
corrections. These should be investigated in more detail. 

Of course, it would be fascinating to study more exotic situations 
using the present approach. In particular one can climb up Seiberg's 
flavor ladder \cite{Seiberg} 
encountering $N_f=N_c$ with baryonic composite fields and 
for $N_f>N_c$ eventually encountering non-Abelian dualities. Breaking 
 $N=2$ supersymmetric gauge theories down to $N=1$ and ordinary gauge 
theories is clearly also of great interest.

\acknowledgments

We are happy to thank Masayasu Harada for many helpful discussions. 
 One of us (F.S.) would like to thank Lucia Pappalardo for helpful 
discussions. 
This work has been supported in part by the US DOE under contract
DE-FG-02-85ER 40231.

\appendix

\section{The $\theta$ dependence}
\label{app:a}
Here we discuss the $\theta$ dependence of the potential Eq.~(\ref{YMTheta}), 
which represents a 
model in which the super-partners have been integrated out of 
the VY Lagrangian. 
To get an indication, which should be accurate for small $\hat{\theta}$, of 
what is happening we eliminate ${\rm Im}F$ from Eq.~(\ref{YMTheta})  by 
solving $\displaystyle{\frac{\partial V}{\partial {\rm Im}F}=0}$ to yield 
\begin{equation}
\Phi = \frac{\gamma \hat{\theta}}{N_c} \ ,
\label{eqofm1}
\end{equation}
where $\Phi$ is defined from $F=-|F| e^{i\Phi}$. 
Substituting Eq.~(\ref{eqofm1}) back into Eq.~(\ref{YMTheta}) together 
with ${\rm Im}F=\tan \Phi \,{\rm Re} F$ gives
\begin{equation}
V=-\frac{2N_c}{\gamma} {\rm Re} F \, \rm{ln}\, 
\left[\frac{N_c}{e \gamma \Lambda^4_{YM} }
\frac{|{\rm Re} F|}{|\cos\left(\frac{\gamma \hat{\theta}}
{N_c}\right)|}\right] \ .
\label{pota1}
\end{equation}
This has a minimum at 
\begin{equation}
\langle |{\rm Re} F|\rangle = \frac{\gamma \Lambda^4_{YM}} 
{N_c} \left| \cos\left[
 \frac{\gamma}{N_c}\left(\theta + 2\pi k\right)\right]\right| \ .
\end{equation}
At the minimum $\displaystyle{\langle V \rangle = 
-\frac{2N_c}{\gamma}\langle |{\rm Re} F|\rangle}$. Thus if $\theta$ is 
very small the global minimum will occur for $k=0$, if 
$\theta=2\pi + $ (very small)  the global minimum will occur for 
$k=-1$, etc. In this way the $N_c-$fold degeneracy of the super 
Yang-Mills theory is broken. 

In order to give a more accurate treatment when $\hat{\theta}$ is 
large, we may recognize that the mass diagonal (evaluated at the 
critical point of  $V$) states are ${\rm Re} F^{\prime}$ and 
${\rm Im}F^{\prime}$ 
where 
\begin{equation}
F^{\prime}=e^{-i\frac{\gamma \hat{\theta}}{2N_c}} F \ .
\label{eigenstate}
\end{equation} 
${\rm Im}F^{\prime}$ has negative squared mass and should, according 
to our earlier discussion, be integrated out from 
$\displaystyle{\frac{\partial V}{\partial ({\rm Im}F^{\prime})}=0}$. 
This has the consequence
\begin{equation}
2\tan\left(\frac{\gamma \hat{\theta}}{2N_c}\right)
\left[l - \rm{ln}\cos \Phi^{\prime}\right]
+2\Phi^{\prime} - \frac{\gamma \hat{\theta}}{N_c}=0
\end{equation} 
where $\displaystyle{l=\rm{ln}\,
\left(\frac{N_c|{\rm Re} F^{\prime}|}{\gamma \Lambda^4_{YM}}\right)}$ 
and 
$\displaystyle{F^{\prime}=-|F^{\prime}|e^{i\Phi^{\prime}}}$.
We may solve for $\Phi^{\prime}$ as a power series in $\hat{\theta}$:
\begin{equation}
\Phi^{\prime}=a\,\hat{\theta} + b\,\hat{\theta}^2 + c\,\hat{\theta}^3 + 
\cdots \ ,
\label{perturbphi}
\end{equation}
with $\displaystyle{a=\frac{\gamma}{2N_c}\left(1 - l\right)}$, $b=0$, 
and 
$\displaystyle{c=-\frac{1}{2}\left(\frac{\gamma}{2N_c}\right)^3
\left(1 - \frac{4}{3}l + l^2\right)}$. Equation (\ref{perturbphi}) may be 
substituted back into $V$ and ${\rm Im}F^{\prime} = \tan \Phi^{\prime}\, 
{\rm Re} F^{\prime}$ eliminated. Correct to first order in $\hat{\theta}$ 
we get 
\begin{equation}
V=+\frac{2N_c}{\gamma}|{\rm Re} F^{\prime}|\, \rm{ln}\,
\left(\frac{N_c |{\rm Re} F^{\prime}|}{e\gamma\Lambda^4_{YM}}\right) 
+ \rm{order}\left(\hat{\theta}^2\right) \ ,
\end{equation}
which is in agreement with Eq.~(\ref{pota1}).

\section{Symmetries of Super QCD}
\label{app:b}

At the classical level and in absence of the matter field mass terms,
 the global symmetry group for supersymmetric QCD is 
\begin{equation}
G=SU_L(N_f)\otimes SU_R(N_f) \otimes U_V(1)\otimes U_A(1)\otimes 
\hat{U}_R(1) \ .
\label{Gclassical}
\end{equation}
The presence of the extra axial $R$ symmetry with respect to QCD is related 
to the gluino. At the superfield level the axial transformations 
are:
\begin{equation}
U_A(1): \qquad W^{\alpha}_a \rightarrow W^{\alpha}_a, \qquad 
Q\rightarrow e^{i \alpha} Q, \qquad  \tilde{Q}\rightarrow e^{i \alpha} 
\tilde{Q} \ .
\label{UA}
\end{equation}
This symmetry is broken at the quantum level by the color Adler-Bell-Jackiw 
anomaly and the anomalous variation of the lagrangian is 
\begin{equation}
\delta_{U_A(1)}{\cal L}=N_f \alpha \left(\frac{g^2}{32\pi^2}
\epsilon_{mnrs} F^{mn}_a F^{rs}_a\right) \ .
\label{ua1var}
\end{equation}
The $\hat{U}_R(1)$ transformation may be chosen as follows:
\begin{eqnarray}
\hat{U}_R(1):&W\left(x,\theta\right) \rightarrow & 
e^{i\alpha\frac{3}{2}}
W\left(x,e^{-i\alpha\frac{3}{2}}\theta\right) \ , \nonumber \\
&{}Q\left(x,\theta\right)\rightarrow & 
e^{i\alpha}Q\left(x,e^{-i\alpha\frac{3}{2}}\theta\right) \ , \label{UR}\\ 
&{\rm same~for~}&\tilde{Q}  \ . \nonumber 
\end{eqnarray} 
The anomalous variation of the fundamental lagrangian is 
\begin{equation}
\delta_{\hat{U}_R(1)}{\cal L}=\left(3N_c - N_f\right)\alpha \left(\frac{g^2}{64\pi^2}
\epsilon_{mnrs} F^{mn}_a F^{rs}_a\right) \ .
\label{ur1var}
\end{equation}
It is possible to build an anomaly free transformation ${U}_R(1)$ which is 
a combination of the 
two previous anomalous $U(1)$'s:
\begin{eqnarray}
{U}_R(1):&W\left(x,\theta\right) \rightarrow & 
e^{-i\alpha N_f}
W\left(x,e^{i\alpha N_f}\theta\right) \ , \nonumber \\
&{}Q\left(x,\theta\right)\rightarrow & 
e^{-i\alpha\left(N_f - N_c \right)}Q
\left(x,e^{i\alpha N_f} \theta\right) \ , \label{URfree}\\ 
&{\rm same~for~}&\tilde{Q}  \ . \nonumber
\end{eqnarray} 
In Table~{\ref{table1}} we summarize the axial charges of the relevant 
elementary and composite fields
\begin{table}[htbp]
\begin{center}
\begin{tabular}{c|ccc}
Field  & $U_A(1)$ &$\hat{U}_R(1)$ & ${U}_R(1)$ \\
\hline
 $\lambda$ & $0$ & $+\frac{3}{2}$ &$-N_f$ \\
 $\psi_{Q},~\psi_{\tilde{Q}}$ &$+1$& $-\frac{1}{2}$& $N_c$ \\
 $\phi_{Q},~\phi_{\tilde{Q}}$ &$+1$& $+1$ & $N_c - N_f$ \\
 $S_{\theta=0}=\phi$ &$0$& $+3$ &$-2N_f$ \\
 $S_{\theta^2}=F$ &$0$& $0$ &$0$ \\
 $T_{\theta=0}=t$ &$+2$&$+2$ &$2\left(N_c - N_f\right)$ \\
 $T_{\theta^2}=F_T$ &$+2$&$-1$ &$2N_c$ 
 \end{tabular}
\end{center}
\caption[]{Axial charges for the relevant elementary and composite fields.}
\label{table1}
\end{table}

\end{document}